\documentclass[12pt]{iopart}
\pdfoutput=1
\usepackage{graphicx}% Include figure files
\usepackage{dcolumn}% Align table columns on decimal point
\usepackage{bm}% bold math
\usepackage{url}
\usepackage{amssymb}
 \expandafter\let\csname equation*\endcsname\relax 
  \expandafter\let\csname endequation*\endcsname\relax 
\usepackage{amsmath}

\include{epsf} 

\newcommand{\N}{{\rm I\! N}}
\newcommand{\R}{{\rm I\! R}}

\begin{document}

\title{Weighted-indexed semi-Markov models for modeling financial returns}
\author{Guglielmo D'Amico}  
\address{Dipartimento di Scienze del Farmaco, Facolt\`a di Farmacia, 
Universit\`a "G. D'Annunzio" di Chieti-Pescara,  66013 Chieti, Italy}
\author{Filippo Petroni}
\address{Dipartimento di Scienze Economiche e Aziendali, Facolt\`a di Economia,
Universit\`a degli studi di Cagliari, 09123 Cagliari, Italy}
\bigskip

\date{\today}

\begin{abstract} 
In this paper we propose a new stochastic model based on a generalization of semi-Markov chains to study the high frequency price dynamics of traded stocks. We assume that the financial returns are described by a weighted indexed semi-Markov chain model. We show, through Monte Carlo simulations, that the model is able to reproduce important stylized facts of financial time series as the first passage time distributions and the persistence of volatility. The model is applied to data from Italian and German stock market from 1 January 2007 until the end of December 2010.  
\end{abstract}
\maketitle

%\indent {\bf Keywords:} first passage time distribution; autocorrelation function; exponentially weighted moving average; Monte Carlo. 
%\\

%\indent {\bf Mathematical Subject Classification 2000:} .

%%%%%%%%%%%%%%%%%%%%%%%%%INTRODUCTION%%%%%%%%%%%%%%%%%%%%%%%%%%%%
\section{Introduction}
Semi-Markov processes (SMP) are a wide class of stochastic processes which generalize at the same time both Markov chains and renewal processes. The main advantage of SMP is that they allow the use of whatever type of waiting time distribution for modeling the time to have a transition from one state to another one. On the contrary, Markovian models have constraints on the distribution of the waiting times in the states which should be necessarily represented by memoryless distributions (exponential or geometric for continuous and discrete time cases respectively). This major flexibility has a price to pay: the parameters to be estimated are more numerous.\\
\indent Semi-Markov processes (SMP) generalizes also non-Markovian models based on continuous time random walks extensively used in the econophysics community, see for example \cite{mai00,rab02}.   
SMP have been used to analyze financial data and to describe different problems ranging from credit rating data modeling \cite{dam05} to the pricing of options \cite{dam09,sil04}.

With the financial industry becoming fully computerized, the amount of recorded data, from daily close all the way down to tick-by-tick level, has exploded. Nowadays, such tick-by-tick high-frequency data are readily available for practitioners and researchers alike \cite{gui97,pet03}. It seemed then natural to us trying to verify the semi-Markov hypothesis of returns on high-frequency data, see \cite{dami11b}. In \cite{dami11b} we proposed a semi-Markov model showing its ability to reproduce some stylized empirical facts such for example the absence of autocorrelations in returns and the gain/loss asymmetry. In that paper we showed also that the autocorrelation in the square of returns is higher with respect to the Markov model. Unfortunately this autocorrelation was still too small compared to the empirical one.  In order to overcome the problem of low autocorrelation, in another paper \cite{dami11a} we proposed an indexed semi-Markov model for price return. More precisely 
 we assumed that the intraday returns (up to one minute frequency) are described by a discrete time homogeneous semi-Markov process where we introduced a memory index which takes into account the periods of different volatility in the market. It is well known that the market volatility is autocorrelated, then periods of high (low) volatility may persist for long time. We made the hypothesis that the kernel of the semi-Markov process depend on which level of volatility the market is at that time. It is to be remarked that the weighted memory index is a stochastic process which do depend on the same Markov Renewal Chain $(J_{n}, T_{n})$ to which the semi-Markov chain is associated. Then, in our model, the high autocorrelation is obtained endogenously without introducing external or latent auxiliary stochastic processes.
To improve further our previous results, in this work, we propose an exponentially weighted index which will be described in the following. 

The database used for the analysis is made of high frequency tick-by-tick price data from all the stock in Italian and German stock market from first of January 2007 until end of December 2010. From prices we then define returns at one minute frequency.

The plan of the paper is as follows. In Section 2 we define the weighted indexed semi-Markov chain model with memory and we explain how to perform a Monte Carlo simulation of its trajectory.  In Section 3, we present the empirical results deriving from the application of our model to real stock market data. Finally, in Section 4 we present our conclusion.

\section{The Weighted-Indexed Semi-Markov Model}
In this section we propose a generalization of the semi-Markov process that is able to represent higher-order dependencies between successive observations of a state variable. One way to increase the memory of the process is by using high-order semi-Markov processes as defined in \cite{limn03} and more recently reviseted and extended in a discrete time framework in \cite{dapefla1}. A more parsimonious model has been defined by \cite{dami11c} and it is showed that it describes appropriately important empirical regularities of financial time series. In this paper we propose a further improvement of the indexed semi-Markov chain model proposed in reference \cite{dami11c} named Weighted-Indexed Semi-Markov Chain (WISMC) model which allows the possibility of reproducing long-term dependence in the square of stock returns in a very efficient way.

\indent  Let us assume that the value of the financial asset under study is described by the time varying asset price $S(t)$. The return at time $t$ calculated over a time interval of lenght $1$ is defined as $\frac{S(t+1)-S(t)}{S(t)}$. The return process changes value in time, then we denote by $\{J_{n}\}_{n\in \N}$ the stochastic process with finite state space $E=\{1,2,...,s\}$ and describing the value of the return process at the $n$-th transition.\\
\indent Let us consider the stochastic process $\{T_{n}\}_{n\in \N}$ with values in $\N$. The random variable $T_{n}$ describes the time in which the $n$-th transition of the price return process occurs.\\
\indent Let us consider also the stochastic process $\{U_{n}^{\lambda}\}_{n\in \N}$ with values in $\R$. The random variable $U_{n}^{\lambda}$ describes the value of the index process at the $n$-th transition.\\
\indent In reference \cite{dami11a} the process $\{U_{n}\}$ was defined as a reward accumulation process linked to the Markov Renewal Process $\{J_{n},T_{n}\}$; in \cite{dami11a} the process $\{U_{n}\}$ was defined as a moving average of the reward process. Here, motivated by the application to financial returns, we consider a more flexible index process defined as follows:
\begin{equation}
\label{funcrela}
U_{n}^{\lambda}=\sum_{k=0}^{n-1}\sum_{a=T_{n-1-k}}^{T_{n-k}-1}f(J_{n-1-k},a,\lambda),
\end{equation}
where $f:E\times \N \times \R \rightarrow \R$ is a Borel measurable bounded function and $U_{0}^{\lambda}$ is known and non-random.\\
\indent The process $U_{n}^{\lambda}$ can be interpreted as an accumulated reward process with the function $f$ as a measure of the weighted rate of reward per unit time. The function $f$ depends on the current time $a$, on the state $J_{n-1-k}$ visited at current time and on the parameter $\lambda$ that represents the weight.\\
In next section a specific functional form of $f$ will be selected in order to produce a real data application.\\
\indent To construct the WISMC model we have to specify a dependence structure between the variables. Toward this end we adopt the following assumption:
\begin{equation}
\label{kernel}
\begin{aligned}
& \mathbb{P}[J_{n+1}=j,\: T_{n+1}-T_{n}\leq t |\sigma(J_{h},T_{h},U_{h}^{\lambda}),\, h=0,...,n, J_{n}=i, U_{n}^{\lambda}=v]\\
& =\mathbb{P}[J_{n+1}=j,\: T_{n+1}-T_{n}\leq t |J_{n}=i, U_{n}^{\lambda}=v]:=Q_{ij}^{\lambda}(v;t),
\end{aligned}
\end{equation}
\noindent where $\sigma(J_{h},T_{h},U_{h}^{\lambda}),\, h\leq n$ is the natural filtration of the three-variate process.\\
\indent The matrix of functions ${\bf Q}^{\lambda}(v;t)=(Q_{ij}^{\lambda}(v;t))_{i,j\in E}$ has a fundamental role in the theory we are going to expose, in recognition of its importance, we call it $\emph{weighted-indexed}$ $\emph{semi-Markov}$ $\emph{kernel}$.\\
\indent The joint process $(J_{n},T_{n})$ depends on the process $U_{n}^{\lambda}$, the latter acts as a stochastic index. Moreover, the index process $U_{n}^{\lambda}$ depends on $(J_{n},T_{n})$ through the functional relationship $(\ref{funcrela})$.\\
\indent Observe that if 
\[
\mathbb{P}[J_{n+1}=j,\: T_{n+1}-T_{n}\leq t |J_{n}=i, U_{n}^{\lambda}=v]=\mathbb{P}[J_{n+1}=j,\: T_{n+1}-T_{n}\leq t |J_{n}=i]
\]
\noindent for all values $v\in \R$ of the index process, then the weigthed indexed semi-Markov kernel degenerates in an ordinary semi-Markov kernel and the WISMC model becomes equivalent to classical semi-Markov chain model as presented for example in \cite{jans06} and \cite{barb08}.\\
\indent The triple of processes $\{J_{n}, T_{n}, U_{n}^{\lambda}\}$ describes the behaviour of the system only in correspondence of the transition times $T_{n}$. To describe the behavior of our model at whatever time $t$ which can be a transition time or a waiting time, we need to define additional stochastic processes.\\
\indent Given the three-dimensional process $\{J_{n}, T_{n}, U_{n}^{\lambda}\}$ and the weighted indexed semi-Markov kernel ${\bf Q}^{\lambda}(v;t)$, we define by
\begin{equation}
\label{stocproc}
\begin{aligned}
& N(t)=\sup\{n\in \mathbb{N}: T_{n}\leq t\};\\
& Z(t)=J_{N(t)};\\
& U^{\lambda}(t)=\sum_{k=0}^{N(t)-1+\theta}\,\,\sum_{a=T_{N(t)+\theta -1-k}}^{(t\wedge T_{N(t)+\theta-k})-1}f(J_{N(t)+\theta-1-k},a,\lambda),
\end{aligned}
\end{equation}
where $\theta =1_{\{t>T_{N(t)}\}}$.\\
\indent The stochastic processes defined in $(\ref{stocproc})$ represent the number of transitions up to time $t$, the state of the system (price return) at time $t$ and the value of the index process (weighted moving average of function of price return) up to $t$, respectively. We refer to $Z(t)$ as a weighted indexed semi-Markov process.\\
\indent The process $U^{\lambda}(t)$ is a generalization of the process $U_{n}^{\lambda}$ where time $t$ can be a transition or a waiting time. It is simple to realize that if $t=T_{n}$ we have that $U^{\lambda}(t)=U_{n}^{\lambda}$.\\  
\indent Let 
$$
p_{ij}^{\lambda}(v):= \mathbb{P}[J_{n+1}=j|J_{n}=i,U_{n}^{\lambda}=v],
$$ 
be the transition probabilities of the embedded indexed Markov chain. It denotes the probability that the next transition is in state $j$ given that at current time the process entered in state $i$ and the index process is equal to $v$. It is simple to realize that
\begin{equation}
p_{ij}^{\lambda}(v)=\lim_{t\rightarrow \infty}Q_{ij}^{\lambda}(v;t).
\end{equation}
\indent Let $H_{i}^{\lambda}(v;\cdot)$ be the sojourn time cumulative distribution in state $i\in E$:
\begin{equation}
H_{i}^{\lambda}(v;t):= \mathbb{P}[ T_{n+1}-T_{n} \leq t |  J_n=i,\, U_{n}^{\lambda}=v ]= \sum_{j\in E}Q_{ij}^{\lambda}(v;t).
\end{equation}
\indent It expresses the probability to make a transition from state $i$ with sojourn time less or equal to $t$ given the indexed process is $v$.\\  \indent The conditional waiting time distribution function $G$ expresses the following probability:
\begin{equation}
\label{G}
G_{ij}^{\lambda}(v;t):=\mathbb{P}[T_{n+1}-T_{n}\leq t \mid J_{n}=i, J_{n+1}=j,U_{n}^{\lambda}=v].
\end{equation}
\indent It is simple to establish that
\begin{eqnarray}
&&G_{ij}^{\lambda}(v;t)=\left\{
                \begin{array}{cl}
                       \ \frac{Q_{ij}^{\lambda}(v;t)}{p_{ij}^{\lambda}(v)}  &\mbox{if $p_{ij}^{\lambda}(v)\neq 0$}\\
                         1  &\mbox{if $p_{ij}^{\lambda}(v)=0$}.\\
                   \end{array}
             \right.
\end{eqnarray}
\indent In the papers \cite{dami11a} and \cite{dami11b} explicit renewal-type equations were given to describe the probabilistic behavior of the indexed semi-Markov chain. It is possible to derive similar results for the WISMC model but here we prefer to not report these results applied to our model because, in the implementation of the model given in the next section, we follow a Monte Carlo simulation based approach. Monte Carlo methods are very useful for simulating the system behavior and represent a way of generate replicable results.
In the following we give a Monte Carlo algorithm in order to simulate a trajectory of a given WISMC in the time interval $[0, T]$. The algorithm consists in repeated random sampling to compute successive visited states of the random variables $\{J_{0}, J_{1},...\}$, the jump times $\{T_{0},T_{1},...\}$ and the index values $\{U_{0}^{\lambda},U_{1}^{\lambda},... \}$ up to the time $T$.\\
The algorithm consists of 5 steps:\\
1) Set $n=0$, $J_{0}=i$, $T_{0}=0$, $U_{0}^{\lambda}=v$, horizon time$=T$;\\
2) Sample $J$ from $p_{J_{n},\cdot}^{\lambda}(U_{n}^{\lambda})$ and set $J_{n+1}=J(\omega)$;\\
3) Sample $W$ from $G_{J_{n},J_{n+1}}^{\lambda}(U_{n}^{\lambda},\cdot)$ and set $T_{n+1}=T_{n}+W(\omega)$;\\
4) Set $U_{n+1}^{\lambda}=\sum_{k=0}^{n}\sum_{a=T_{n-k}}^{T_{n+1-k}-1}f(J_{n-k},a,\lambda)$;\\
5) If $T_{n+1}\geq T$ stop\\
\indent else set $n=n+1$ and go to 2).

\section{Empirical results}

In the following we show as our model performs comparing its statistical features and those of real data returns.
The comparison is done by means of Monte Carlo simulations according to the algorithms described in the previous section.

For our analysis we choose 4 stocks from two databases of tick-by-tick quotes of real stocks from the Italian Stock Exchange (``Borsa Italiana")
and the German Stock Exchange (``Deutsche B\"orse"). The chosen stocks are ENI and FIAT from the Italian database and Allianz and VolksWagen from the German database.The period used goes from January 2007 to December 2010 (4 full years). 
The data have been re-sampled to have 1 minute frequency. The number of returns analyzed is then roughly $500*10^3$ for each stock.

\begin{figure}
\centering
\includegraphics[width=8cm]{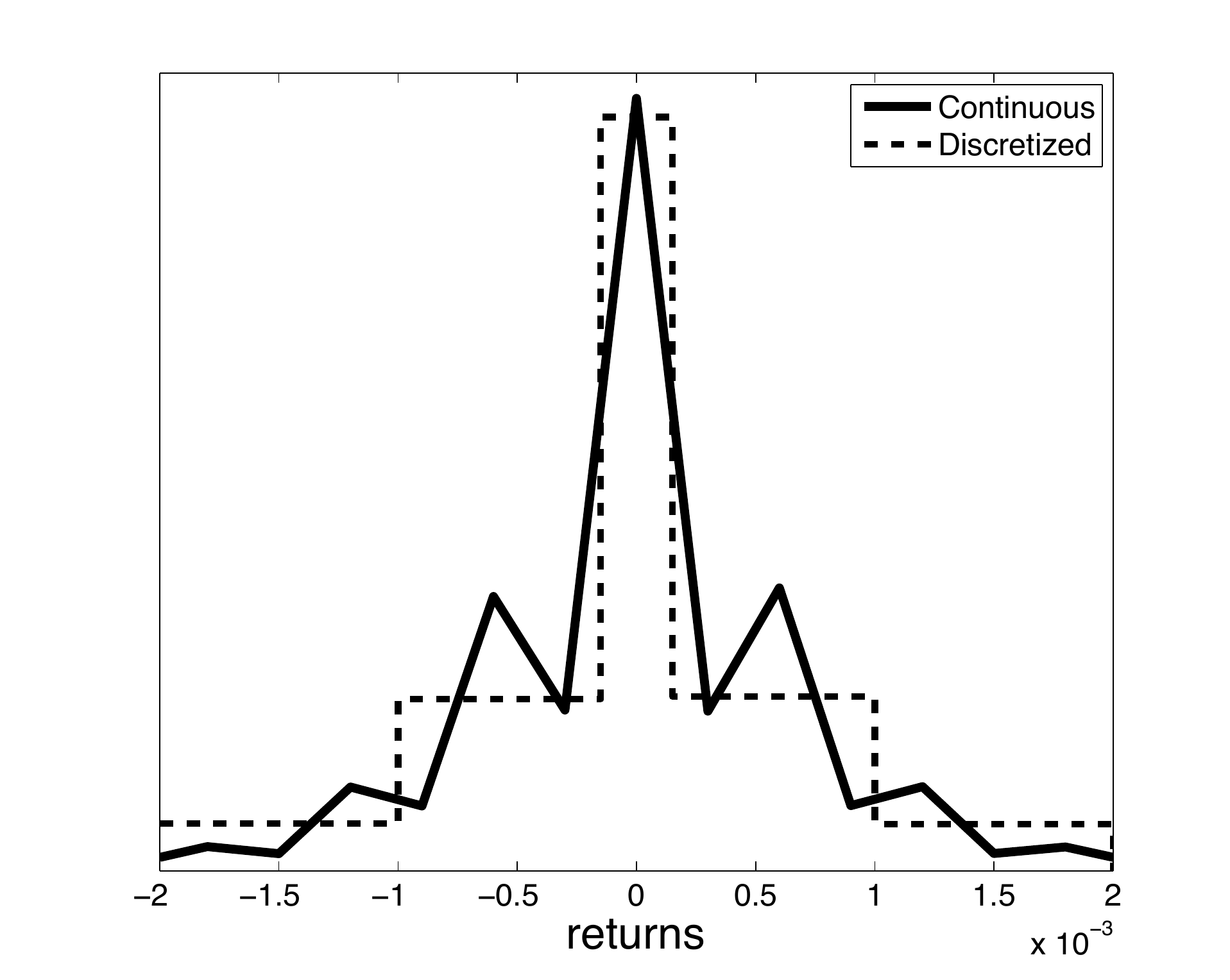}
\caption{Discretization of returns} \label{discR}
\end{figure}

\begin{figure}
\centering
\includegraphics[width=8cm]{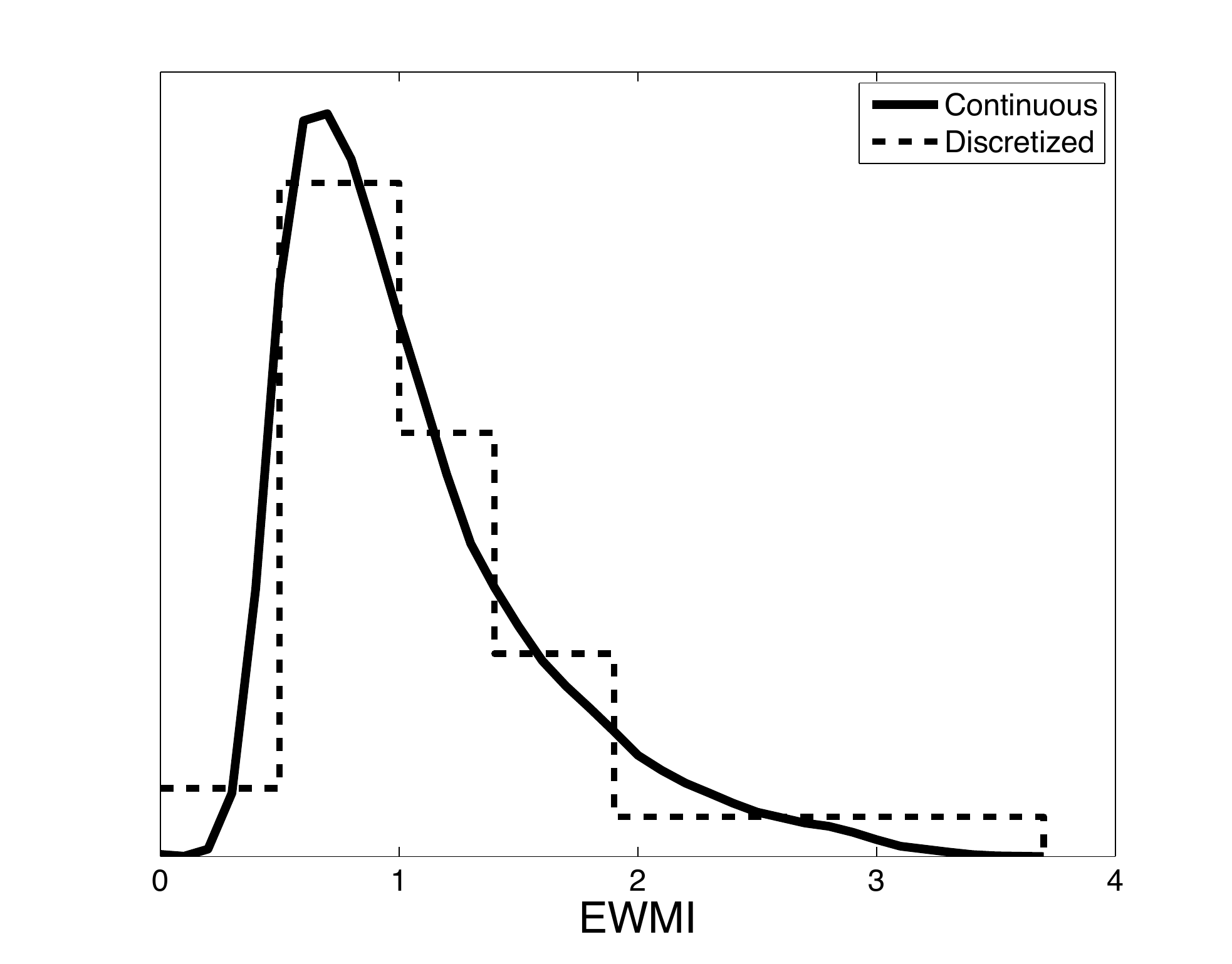}
\caption{Discretization of index values} \label{discU}
\end{figure}

To be able to model returns has a semi-Markov process the state space has been discretized.
In the 4 examples shown in this work, we discretized returns into 5 states chosen to be symmetrical with respect to returns equal zero and to keep the shape of the distribution unchanged. Returns are in fact already discretized in real data due to the discretization of stock prices which is fixed by each stock exchange and depends on the value of the stock. Just to make an example, in the Italian stock market for stocks with value between 5.0001 and 10 euros the minimum variation is fixed to 0.005 euros (usually called tick). We then tried to remain as much as possible close to this discretization. In Figure \ref{discR} we show an example of the discretization of the returns of one of the analyzed stocks. 
The model described in the previous section requires the specification of a function $f$ in the definition of the weighted index $U_{n}^{\lambda}$ in  (\ref{funcrela}). 
Let us briefly remind that volatility of real market is long range positively autocorrelated and then clustered in time. This implies that, in the stock market, there are periods of high and low volatility. Motivated by this empirical facts we suppose that also the transition probabilities depends on whether the market is in a high volatility period or in a low one. In a previous work \cite{dami11a}, for simplicity reason, we used a moving average of the squares of returns as the index variable $U$. In that case we imposed that the index depended only on a memory $m$ which was the number of transitions in the past used for the moving average. In this work we decided to use a more appropriate expression for $f$. We use an exponentially weighted moving average (EWMA) of the squares of returns which as the following expression:
 \begin{equation}
\label{funct}
f(J_{n-1-k},a,\lambda)=\frac{\lambda^{T_{n}-a} J_{n-1-k}^2}{\sum_{k=0}^{n-1}\sum_{a=T_{n-1-k}}^{T_{n-k}-1}\lambda^{T_{n}-a}}
\end{equation}
\noindent and consequently the index process becomes
\begin{equation}
\label{ewma}
U_{n}^{\lambda}=\sum_{k=0}^{n-1}\sum_{a=T_{n-1-k}}^{T_{n-k}-1}\Bigg(\frac{\lambda^{T_{n}-a} J_{n-1-k}^2}{\sum_{k=0}^{n-1}\sum_{a=T_{n-1-k}}^{T_{n-k}-1}\lambda^{T_{n}-a}}\Bigg).
\end{equation}
The index $U^\lambda$ was also discretized into 5 states of low, medium low, medium, medium high and high volatility. An example of the discretization used in the analysis is shown in Figure \ref{discU}.

\begin{figure}
\centering
\includegraphics[height=9cm]{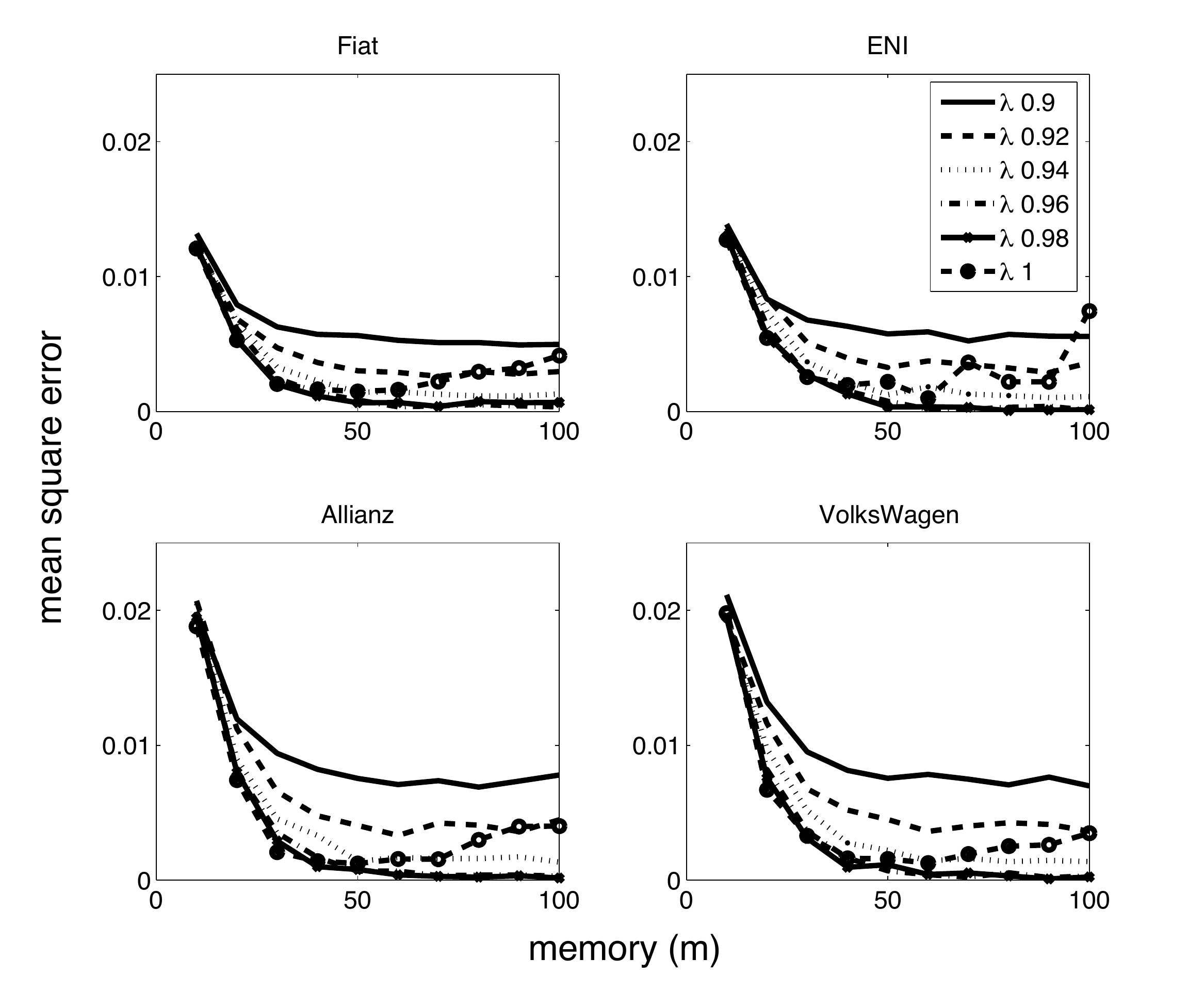}
\caption{Mean square error between autocorrelation functions from real and simulated data as functions of $m$ and for different values of $\lambda$.}\label{fig1}
\end{figure}

\begin{figure}
\centering
\includegraphics[height=9cm]{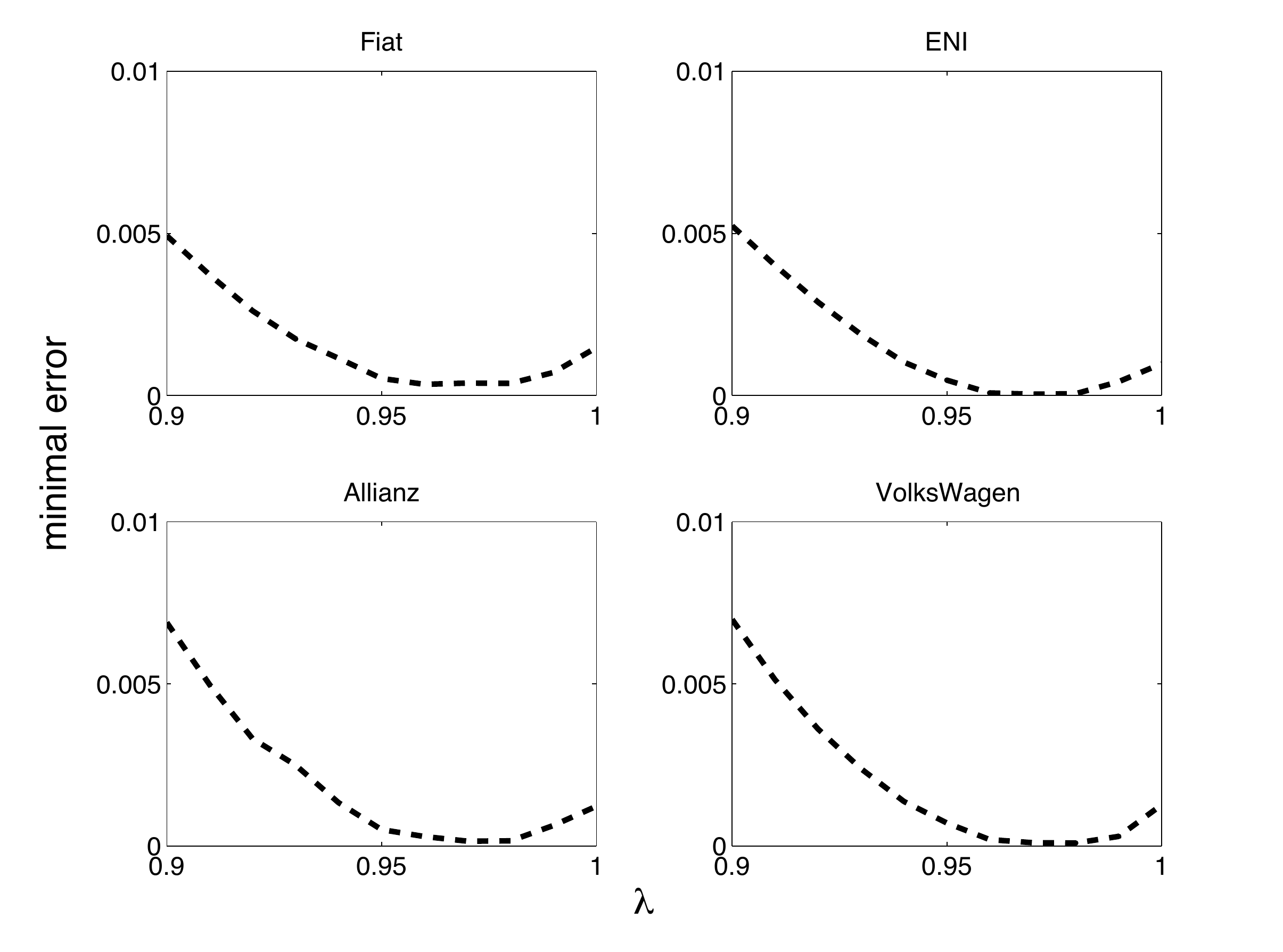}
\caption{Mean square error between autocorrelation functions from real and simulated data as functions of $\lambda$.}\label{fig2}
\end{figure}

\begin{figure}
\centering
\includegraphics[height=8cm]{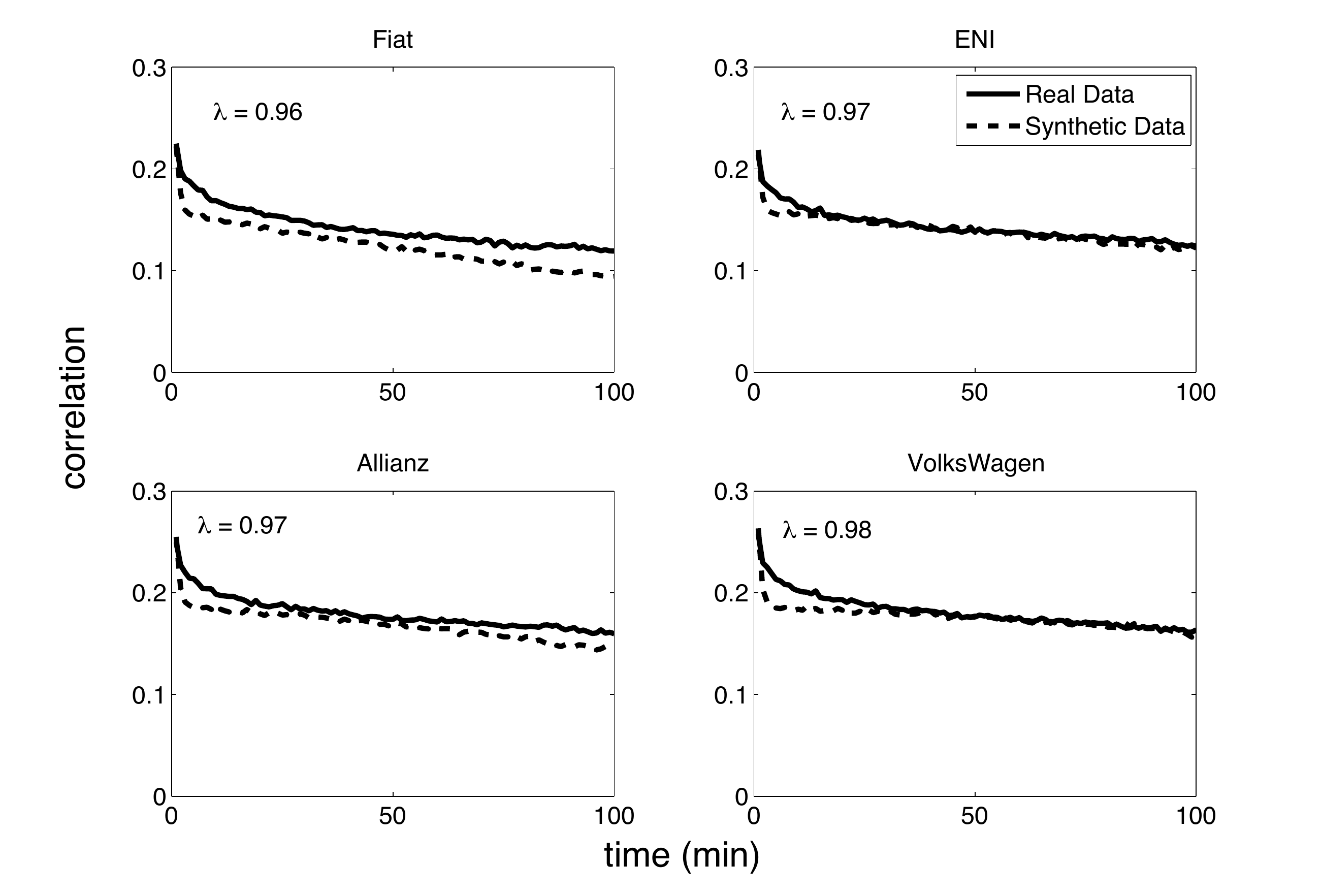}
\caption{Autocorrelation functions of real data (solid line) and synthetic (dashed line) time series for the analyzed stocks.}\label{fig3}
\end{figure}

\begin{figure}
\centering
\includegraphics[height=8cm]{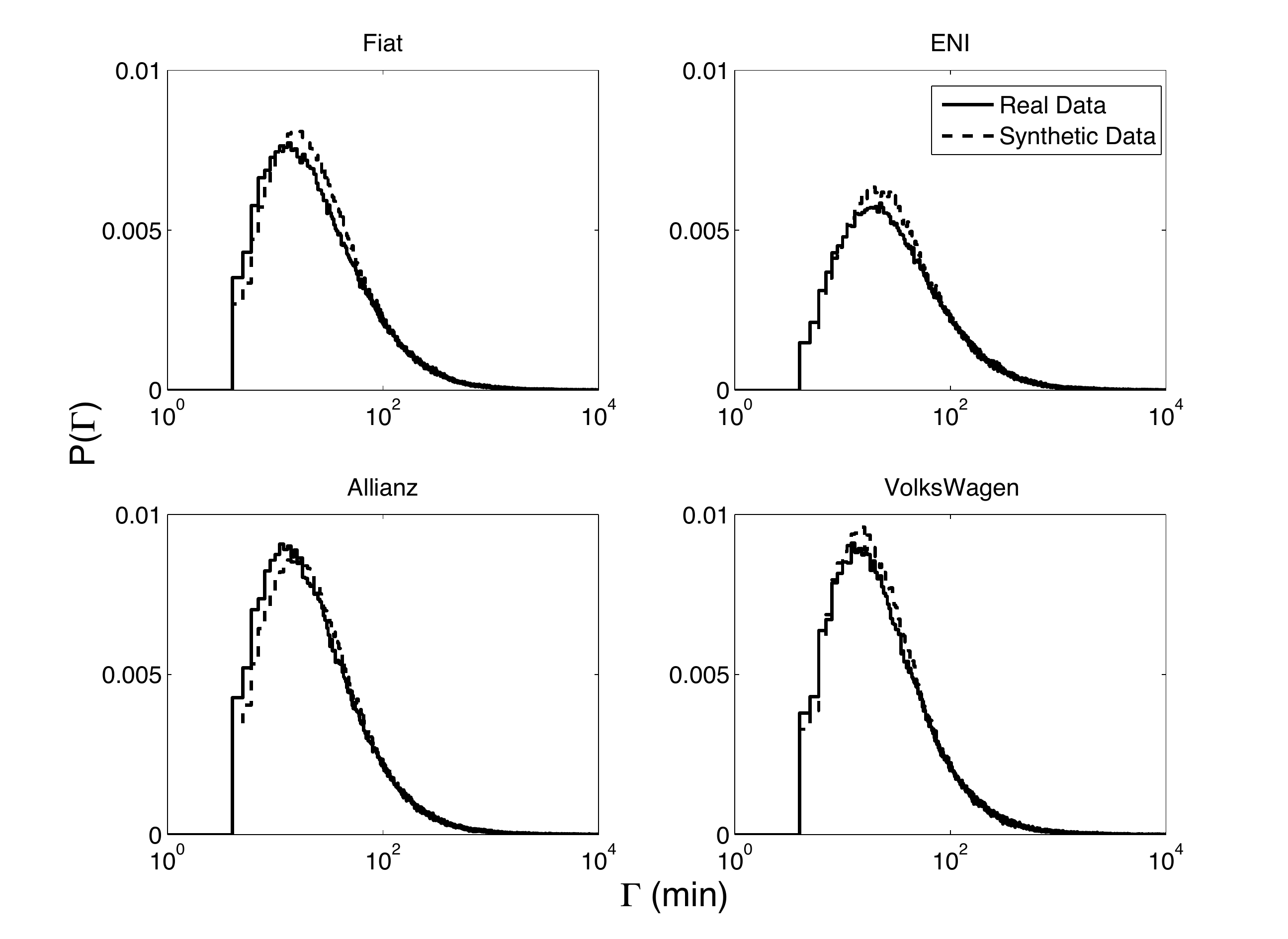}
\caption{First passage time distribution of real data (solid line) and synthetic (dashed line) time series for the analyzed stocks.}\label{fpt}
\end{figure}

A very important feature of stock market data is that, while returns are uncorrelated and show an i.i.d. like behavior, their square or absolute values are long range correlated. It is very important that theoretical models of returns reproduce this features.   
We then tested our model to check whether it is able to reproduce such behavior. 
Given the presence of the parameter $\lambda$ in the index function, we tested the autocorrelation behavior as a function of $\lambda$. Note that in the definition of the index variable the EWMA is performed over all the previous square of returns each with its weight. Before summing over all past returns we decided to check whether a better memory time $m$ exists. For this reason we checked our model also against this other parameter.
With this choice formula $(\ref{ewma})$ takes the form:
\begin{equation}
\label{ewma_m}
U_{n}^{\lambda}(m)=\sum_{k=n-m}^{n-1}\sum_{a=T_{n-1-k}}^{T_{n-k}-1}\Bigg(\frac{\lambda^{T_{n}-a} J_{n-1-k}^2}{\sum_{k=n-m}^{n-1}\sum_{a=T_{n-1-k}}^{T_{n-k}-1}\lambda^{T_{n}-a}}\Bigg).
\end{equation}
We remind the definition of the autocorrelation function: if $Z$ indicates returns, the time lagged $(\tau)$ autocorrelation of the square of returns is defined as 
\begin{equation}
\label{autosquare}
\Sigma(\tau)=\frac{Cov(Z^2(t+\tau),Z^2(t))}{Var(Z^2(t))}
\end{equation}
We estimated $\Sigma(\tau)$ for real data and for returns time series simulated with different values of the memory time $m$ and the weights $\lambda$. 
The time lag $\tau$ was made to run from 1 minute up to 100 minutes. Note that to be able to compare results for $\Sigma(\tau)$ each
simulated time series was generated with the same length as real data.
In Figure \ref{fig1} we show the mean square error between $\Sigma(\tau)$ obtained from real and simulated returns (using definition $(\ref{ewma_m})$ for the index process) for the four stocks analyzed and for different $m$ and  $\lambda$.
Let us make some considerations on the results shown in Figure \ref{fig1}: $m$ should be chosen as big as possible and then definition $(\ref{ewma})$ is appropriate as far as $\lambda$ is chosen less than $1$, in fact, in this last case definition (\ref{ewma}) becomes equivalent to a moving average without weights and results presented in \cite{dami11a} hold for $m$.
In Figure \ref{fig2} we show again the mean square error but only as a function of the weights $\lambda$ then using definition $(\ref{ewma})$ for the index process. 
We can notice that the behavior is very similar for the different analyzed stocks even if the best value for $\lambda$ is not the same for all of them.
As it is possible to see the best values of $\lambda$ for the stocks Fiat, Eni, Allianz and VolksWagen are $0.96$, $0.97$, $0.97$ and $0.98$, respectively.\\
The comparison between the autocorrelations for the best values of $\lambda$ for each stock and real data is shown in Figure \ref{fig3}.
This figure shows that real and synthetic data have almost the same autocorrelation function for the square of returns. \\
We tested our model also to verify if it is able to reproduce the feature shown by real data regarding the first passage time (fpt) distribution  \cite{dami11b,sim02,jen04}. Let us remind here the definition of fpt: 
the fpt for an investment made at time $t$ at price $S(t)$ is defined as the time interval $\tau = t'-t$, $t'>t$ where the relation $S(t+\tau)/S(t) \geq \rho$ is fulfilled for the first time. We will denote the fpt as $\Gamma_\rho(t)$. Then $$\Gamma_\rho(t)=\min\{\tau \geq 0 ; S(t+\tau)/S(t) \geq \rho\}.$$
In \cite{dami11b} we have shown how to calculate analytically such distribution for a semi-Markov process then we will not repeat that here. 
Using the best values for $\lambda$ for each stock Fiat, Eni, Allianz and VolksWagen and choosing a value $\rho=1.005$ for all of them we compare in Figure \ref{fpt} results for the first passage time distribution for each stock. It can be noticed that they are almost identical improving the results obtained for a simple semi-Markov process presented in \cite{dami11b}.

The results obtained here improve those obtained in our previous work \cite{dami11a,dami11b} even further showing that the semi-Markov approach is adequate to model high frequency financial time series.

\section{Conclusions}
We have modeled financial price changes through a semi-Markov model where we have added a weighted index. Our work is motivated by two main results: the existence in the market of periods of low and high volatility and our previous work \cite{dami11a}, where we showed that an indexed semi-Markov model, is able to capture almost all the correlation in the square of returns present in real data.
The results presented here show that the semi-Markov kernel is influenced by the past volatility and that its influence decreases exponentially with time. In fact, if the past volatility is used as an exponentially weighted index, the model is able to reproduce almost exactly the behavior of market returns: the returns generated by the model are uncorrelated while the square of returns present a long range correlation very similar to that of real data. 

We have also shown, by analyzing different stocks from different markets (Italian and German), that results do not depend on the particular stock chosen for the analysis even if the value of the weights may depends on the stock.

We stress that out model is very different from those of the ARCH/GARCH family. We do not model directly the volatility as a correlated process. We model returns and by considering the semi-Markov kernel conditioned by a weighted index the volatility correlation comes out freely.

%\section*{Acknowledgments}
%We are very grateful to Maurizio Serva, Raimondo Manca and Giuseppe Di Biase for helpful discussion. We also thank an anonymous referee for his precious comments.
%

%
\newpage

\end{document}